\documentclass[12pt, twoside]{article}
\usepackage{a4wide,amssymb,axodraw,epsfig}

\def\de{\partial}

\def\a{\alpha}

\def\b{\beta}

\def\g{\gamma}

\def\G{\Gamma}

\def\d{\delta}

\def\D{\Delta}

\def\e{\eta}

\def\la{\lambda}

\def\La{\Lambda}

\def\k{\kappa}

\def\m{\mu}

\def\n{\nu}

\def\r{\rho}

\def\s{\sigma}

\def\t{\tau}

{\rm }

\def\vf{\varphi}

\def\de{\partial}

\newcommand{\be}{\begin{equation}}

\newcommand{\ee}{\end{equation}}

\newcommand{\bea}{\begin{eqnarray}}

\newcommand{\eea}{\end{eqnarray}}

\newcommand{\beqar}{\begin{eqnarray*}}

\newcommand{\eeqar}{\end{eqnarray*}}

\newcommand{\eg}{{\it e.g.,}\ }

\newcommand{\ie}{{\it i.e.,}\ }

\newcommand{\reef}[1]{(\ref{#1})}

\begin{document}

\begin{titlepage}
%
%

\rightline{July 2007}

\begin{centering}
\vspace{1cm}
{\Large {\bf Induced cosmology on a regularized brane \\in
six-dimensional flux compactification}}\\

\vspace{1.5cm}

 {\bf Eleftherios Papantonopoulos}$^{a,*}$, {\bf Antonios~Papazoglou} $^{b,c,**}$ \\ and  {\bf Vassilios Zamarias}$^{a,***}$\\
\vspace{.2in}

$^{a}$ Department of Physics, National Technical University of
Athens, \\
Zografou Campus GR 157 73, Athens, Greece. \\
\vspace{3mm}
$^{b}$ APC\footnote{UMR 7164(CNRS, Universit\'e Paris 7, CEA, Obervatoire de Paris)}, 10 rue Alice Domon et L\'eonie Duquet, \\
75205 Paris Cedex 13, France. \\
\vspace{3mm}
$^{c}$ GReCO/IAP\footnote{UMR 7095(CNRS, Universit\'e Paris 6)}, 98 bis Boulevard Arago, \\
75014 Paris, France.

\end{centering}
\vspace{2cm}

\begin{abstract}
We consider a six-dimensional Einstein-Maxwell system compactified
in an axisymmetric two-dimensional space with one capped
regularized conical brane of codimension one. We study the
cosmological evolution which is induced on the regularized brane
as it  moves in between known {\it static} bulk and cap solutions.
Looking at the resulting Friedmann equation, we see that the brane
cosmology at high energies is dominated by a five-dimensional
$\rho^2$ energy density term. At low energies, we obtain a
 Friedmann equation with a term linear to the energy density
 with, however, negative coefficient in the small four-brane radius
 limit (\ie with negative effective Newton's constant).
 We discuss ways out of this problem.

\end{abstract}

\vspace{1.5cm}
\begin{flushleft}

$^{*}~$ e-mail address: lpapa@central.ntua.gr \\
$ ^{**}$ e-mail address: papazogl@iap.fr\\
$ ^{***}$ e-mail address: zamarias@central.ntua.gr

\end{flushleft}
\end{titlepage}

\section{Introduction}

Six-dimensional brane world models are rather interesting to be
studied for a number of reasons. Firstly, it was in the framework
of brane theories with two extra transverse dimensions that the
large (sub-millimeter) extra dimensions proposal for the
resolution of the electroweak hierarchy problem was provided
\cite{AADD}.  This scenario makes the study of extra dimensional
theories, and string theory in particular, relevant to low energy
phenomenology (colliders as well as astrophysical and cosmological
observations) and is testable in the very near future. The
different possibilities of realizing a  brane world model in six
dimensions must therefore be studied to allow for a comparison
with experiment.

Quite a different motivation has been the proposal to ameliorate
the cosmological constant problem, using codimension-2 branes (for
a recent review on the subject see \cite{koz}). These branes have
the interesting property that their vacuum energy instead of
curving their world-volume, just introduces a deficit angle in the
local geometry \cite{CLP}. Models with this property which exhibit
no fine-tuning between the brane and bulk quantities have been
known as self-tuning (for early attempts to find similar models in
five dimensions see \cite{5d}). Such self-tuning models with flux
compactification \cite{6dflux,susy} have been extensively looked,
but the flux quantization condition always introduces a fine
tuning \cite{fluxquant}, unless one allows for singularities more
severe than conical \cite{noncon}. Alternative sigma-model
compactifications have been shown to satisfy the self-tuning
requirements \cite{6dsigma}.  However, the successful resolution
of the cosmological constant problem would also require that there
are  no fine-tuning between bulk parameters themselves. No such
self-tuning model has been found yet with all these properties.

A further motivation in studying such models with codimension-2
branes is that gravity on them is purely understood. The
introduction of matter (\ie anything different from vacuum energy)
on them, immediately introduces malicious non-conical
singularities \cite{Cline}. A way out of this problem is to
complicate the gravity dynamics by adding a Gauss-Bonnet term in
the bulk or an induced curvature term on the brane, in which case
the singularity structure of the theory is altered and non-trivial
matter is allowed \cite{GB}. However, the components of the
energy-momentum tensor of the brane and the bulk are tuned
artificially and the brane matter is rather restricted
\cite{GBcon}. Alternatively, one can regularize the codimension-2
branes by introducing some thickness and then consider matter on
them \cite{regular}. For example,  one can  mimic the brane by a
six-dimensional vortex (as \eg in \cite{uzpe}), a procedure which
becomes a rather difficult task if matter is added on it.

Another way of regularization was proposed recently, which
consists merely of the reduction of codimensionallity of the
brane. In this approach, the bulk around the codimension-2 brane
is cut close to the conical tip and it is replaced by a
codimension-1 brane which is capped by appropriate bulk sections
\cite{PST} (see \cite{Gott} for a similar regularization of cosmic
strings in flat spacetime). This regularization has been applied
to flux compactification systems in six dimensions for unwarped
``rugby-ball''-like solutions in \cite{PST}, for warped solutions
with conical branes (with or without supersymmetry) in \cite{ppz}
and for even more general warped solutions allowing non-conical
branes in \cite{tas}. Specific brane energy-momentum tensor is
required to build static solutions, involving a brane field with
Goldstone-like dynamics. It is interesting to note that in the
non-supersymmetric case, only quantized warpings are allowed
\cite{ppz}.

In the present paper we will try to go a step further and consider
an isotropic cosmological fluid on the above-mentioned regularized
branes. To have a cosmological evolution on the regularized
branes, the brane world-volume should be expanding and in general
 the bulk space should also evolve in time. Instead of tackling this problem in its full
generality, which seems a formidable task, in the present work we
will consider the motion of the regularized codimension-1 brane in
the space between the bulk and the brane-cap which remains static
(see \eg \cite{Cuadros-Melgar:2005ex}).  In this way, a
cosmological evolution will be induced on the brane in a similar
way as in the
  mirage
cosmology \cite{mirage}, but with the inclusion of the
back-reaction of the brane energy density (\ie the brane is not
considered merely a probe one). Since in the mirage cosmology, the
four-dimensional scale factor descends from the warp factor in the
four-dimensional part of the bulk metric, we will discuss the
regularized brane in the case of warped bulk \cite{ppz}, rather
than unwarped bulk \cite{PST}. It is worth noting that the above
procedure provided in five dimensions the most general isotropic
brane cosmological solutions \cite{5dcosmo}.

Solving the Israel junction conditions,  which play the r\^ole of
the equations of motion of the codimension-1 brane, we find the
Friedmann
 equation on the brane which at early times is dominated by an energy density term
 proportional to $\rho^2$,  like in the Rundall-Sundrum model in five dimensions.
  To recover four-dimensional cosmology at late times, we split as usual the
  energy-momentum tensor to a part
  which is the contribution of the static vacuum brane and to a part of additional matter.
    We find a regime which has a four-dimensional
   dependence on the energy density. However, in the interesting
    case where the brane moves close to its equilibrium point,
     which in turn is close to the would-be conical singularity,
      the coefficient of the linear to the energy density term is negative (\ie we obtain negative
   effective   Newton's  constant).
       Thus, we cannot recover the standard cosmology at late times. This seems
to be the consequence of considering the bulk sections static. It
is possible, that this behaviour is due to a ghost mode appearing
among the perturbations of the system, after imposing the
staticity of the bulk sections. Furthermore, the above result
points out that there is a difference  between the six-dimensional
brane cosmology in comparison to  the five-dimensional one. The
study of brane cosmology in Einstein gravity in five dimensions,
can be made either in a gauge where the bulk is time-dependent and
the brane lies at a fixed position, or in a gauge where the bulk
is static and the brane  movement into the bulk induces a
cosmological evolution on it \cite{equivalence}. This, however,
does not seem to hold in six-dimensions anymore.

The paper is organized as follows. In Sec.~2 we review the static
regularized brane solution in a bulk of general warping. In Sec.~3
we derive the equations of motion of the codimension-1 brane and
in Sec.~4 we study the induced cosmological evolution on the
moving brane. Finally in Sec.~5 we draw our conclusions.

\section{Setup and static brane solutions}

\begin{figure}[t]
\begin{center}
\begin{picture}(200,160)(-60,-20)

\Text(45,155)[c]{$r=+1$}
\Text(45,-10)[c]{$r=-1$}

\Text(125,50)[c]{Bulk: $c_0,~R_0,~\a$}

\Text(125,130)[c]{Cap: $c_c,~R_c,~\a$}

\Text(-65,142)[c]{4-brane}
\Text(-65,129)[c]{$\la,~v$}
\Text(-65,114)[c]{$r=r_c$}

\BBoxc(-65,-5.5)(50,30)

\Text(-65,2)[c]{3-brane}
\Text(-65,-13)[c]{$T$}

\Vertex(42,2){2}

\Oval(-65,128)(22,38)(0)

\LongArrow(-20,125)(25,125)
\LongArrow(-30,2)(33,2)

\epsfig{file=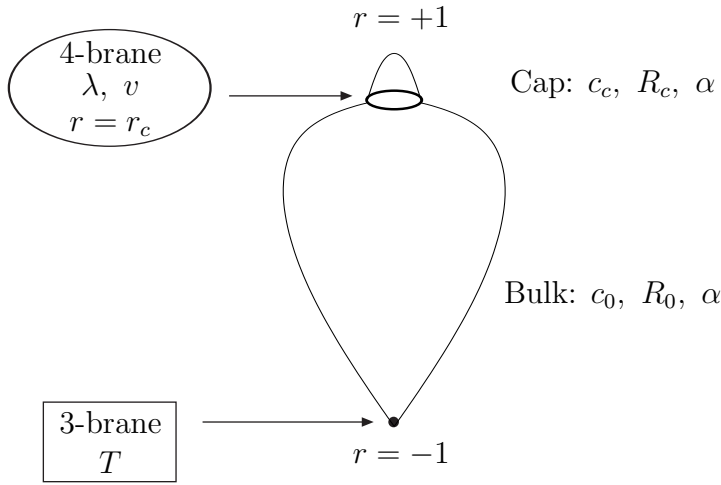,width=3cm,height=5cm}

\end{picture}
\caption{ The internal space where the upper codimension-2 singularity
has been  regularized with the introduction of a ring-like
codimension-1 brane. The parameters of the action and the solution are denoted
in the appropriate part of the internal space.}
\label{internalfig}
\end{center}
\end{figure}

Before discussing the time-dependent scenario, let us remind
ourselves  of the static solution which we will use in the
following for the brane motion. The bulk theory that we will use
is a six-dimensional Einstein-Maxwell system  which in the
presence of a positive cosmological constant and a gauge flux,
spontaneously compactifies the internal space \cite{spontan}.  The
known axisymmetric solutions have in general two codimension-2
singularities at the poles of a deformed sphere \cite{japs}. We
will study the case where only one (\eg the upper) codimension-2
brane is regularized by the introduction of a ring-like brane  at
$r=r_c$ with an appropriate cap. The dynamics of the system is
given by the following action \be S= \int d^6 x \sqrt{-g} \left(
{M^4 \over 2} R - \La_i -{1 \over 4}{\cal F}^2 \right) - \int d^5
x \sqrt{-\g_+}\left(\la +{v^2 \over 2} (\tilde{D}_{\hat{\m}} \s)^2
\right)-\int d^4x \sqrt{-\g_-} ~T~, \ee where $M$ is the
six-dimensional  fundamental Planck mass, $\La_i$ are the bulk
($i=0$) and cap ($i=c$) cosmological constants, ${\cal F}_{MN}$
the gauge field strength, $T$ the tension of the lower
codimension-2 brane, $\la$ the 4-brane tension, $\s$ the 4-brane
Goldstone scalar field necessary for the regularization and $v$
the vev of the Higgs field from which the Goldstone field
originates. For the coupling between the Goldstone field and the
bulk gauge field we use the notation    $\tilde{D}_{\hat{\m}} \s =
\de_{\hat{\m}} \s -e~ a_{\hat{\m}}$, with $a_{\hat{\m}}={\cal A}_M
\de_{\hat{\m}} X^M$ the pullback of the  gauge field on the
ring-like brane and $e$ its coupling to the scalar field. In  the
above action we omitted the Gibbons-Hawking term. The
configuration is shown in more detail in Fig.\ref{internalfig}.

The solution for the bulk and cap regions depends on a parameter
$\a$ which is a measure of the warping of the space (for $\a=1$ we
obtain the unwarped case) and is given by \cite{ppz} \bea
ds_6^2&=&  z^2 \eta_{\m\n} dx^\m dx^\n + R_i^2\left[ {dr^2 \over f} +c_i^2 f~ d\vf^2 \right]~,\\
{\cal F}_{r \vf}&=&  -c_i R_i M^2 S \cdot {1 \over z^4}~, \eea
with $R_i^2=M^4/(2 \La_i)$ and the following bulk functions \bea
z(r)&=&{1 \over 2}[(1-\a)r+(1+\a)]\\ f(r)&=& {1 \over 5
(1-\a)^2}\left[ -z^2 + {1-\a^8 \over 1-\a^3}\cdot{1 \over z^3
}-\a^3{1-\a^5 \over 1-\a^3}\cdot{1 \over z^6} \right]~, \eea with
$S(\a)= \sqrt{{3 \over 5}\a^3{1-\a^5 \over 1-\a^3}}$. The range of
the internal space coordinates is $-1 \leq r \leq 1$ and $0\leq
\vf <2 \pi$. Taking into account that in the limit $r \to \pm 1$,
it is $f \to 2(1\mp r)X_\pm$ with the constants $X_\pm$ given by
\be X_+= {5+3\a^8-8\a^3 \over 20 (1-\a)(1-\a^3)} ~~~ ,~~~ X_-=
{3+5\a^8-8\a^5 \over 20 \a^4(1-\a)(1-\a^3)} ~, \ee the cap is
smooth at $r=+1$ as long as is $c_c=1/X_+$. Furthermore, the
metric is continuous if $c_0 R_0=c_c R_c$, which gives $R_c= \b_+
R_0$ with $\b_+=X_+ c_0$ \footnote{In this brief presentation of
the background, we have taken $\xi=1$ in comparison with
\cite{ppz}. The physical quantities, however, are
$\xi$-independent and depend only on $\b_+$.}. The conical
singularity at $r=-1$ is supported by a codimension-2 brane with
tension \be T=2\pi M^4 (1- c_0 X_- ) \ee while the parameters of
the $4$-brane $\la$, $v$ are fixed by the radii $R_0$, $R_c$ and
the brane position $r_c$ \cite{ppz}. Furthermore, the gauge field
is quantized as \be 2 c_0 R_0 M^2 e~Y = N   ~~~,~~~N \in
\mathbb{Z}~, \label{N} \ee with $Y={(1-\a^3) \over 3 \a^3
(1-\a)}~S$ and the brane scalar field has  solution $\s= n \vf$
with $n \in \mathbb{Z}$. The two quantum numbers $n$, $N$ are
related through the junction conditions as
 \be n={N
\over 2}{2 \over (1-\a^3)}\left[{5(1-\a^8) \over 8 (1-\a^5)}
-\a^3\right]~. \ee
 Since the quantities $n$, $N$ are integers,
the above relation imposes
 a restriction to the values of the admissible warpings $\a$, which implies that static solutions are
consistent only for discrete values of the warping $\a$.

\section{Moving brane junction conditions}

To study the cosmological evolution on the 4-brane we introduce an
energy momentum tensor of a perfect fluid on the brane. Then the
total energy momentum tensor $t_{\hat{\m}
\hat{\n}}^{(br)}=-(2/\sqrt{-\g_+}) \d S_{br}/ \d \g_+^{\m\n}$
(where $S_{br}$ is the brane action)  will be given by \be
t_{\hat{\m}}^{\hat{\n} ~(br)}={\rm diag}(-\r, P,P,P,\hat{P}) \ee
and a possible coupling of the brane matter to the bulk gauge
field (consistent with the cosmological symmetries) by \be \d
S_{br} / \d a^{\hat{\k}}=(l,L,L,L,\hat{L}) \ee Splitting the above
quantities to one part responsible for the static solution and
another expressing the presence of matter on the brane, we have
 \bea
\r&=&\la + {v^2 (n-e {\cal A}_\vf^+)^2 \over 2 c_0^2 R_{0}^2  f(r_c)} + \r_m \equiv  \r_0+ \r_m \label{rsplit} \\
P&=&-\la - {v^2 (n-e {\cal A}_\vf^+)^2 \over 2 c_0^2 R_{0}^2  f(r_c)} +P_m \equiv  -\r_0+ P_m\\
\hat{P}&=&-\la + {v^2 (n-e {\cal A}_\vf^+)^2 \over 2 c_0^2 R_{0}^2  f(r_c)} + \hat{P}_m\\
l&=&l_m\\
L&=&L_m \\
\hat{L}&=& e  v^2 (n-e {\cal A}_\vf^+)+\hat{L}_m \eea with all the
other quantities vanishing and $\r_m$, $P_m$, $\hat{P}_m$, $l_m$,
$L_m$, $\hat{L}_m$   the matter contributions.

In general, the inclusion of the above matter contributions will
have  the effect of giving some time dependence {\it both} to the
brane as well as the bulk solutions. Since the study of the full
time dependent problem is rather difficult, we will make in the
present paper the approximation that the bulk remains {\it static}
and that the brane matter merely makes the brane to move between
the two static bulk sections, away from its equilibrium point
$r=r_c$. This is to be regarded as a first step towards
understanding the generic brane cosmological evolution.

To embed  the brane in the static bulk, let us take  the brane
coordinates  be $\s^{\hat{\m}}=(\s,x^i,\vf)$.  [The brane-time
coordinate $\s$ is not to be confused with the Goldstone field
$\s$ which will not appear in our subsequent analysis.] Then the
brane embedding $X^M$ in the bulk is
 taken  for both sections to be
\be
X^i=x^i ~~~ ,~~~ X^r={\cal R}(\s)~~~ {\rm and}  ~~~ X^\vf=\vf~,
\ee
while for the time coordinate embedding we choose for the outer  bulk  section
\be
X^0_{(out)}=\s~,
\ee
 and for the inner cap section
\be
X^0_{(in)}=T(\s)~.
\ee

 The continuity of the induced metric $\g_{\hat{\m} \hat{\n}} = g_{MN} \de_{\hat{\m}} X^M \de_{\hat{\n}} X^N $,
  apart from the relation $c_0 R_0= c_c R_c$ as in the static case,
   gives a relation of the time coordinate $T$ in the upper cap
    region with the brane time coordinate $\s$ (dots are with respect to $\s$)
\be
\dot{T}^2 \left(1 - \b_+^2{\dot{\cal
R}^2 \over \dot{T}^2} {R_0^2 \over f z^2} \right)= \left(1 -
\dot{\cal R}^2 {R_0^2 \over f z^2} \right) ~.\label{T}
\ee
Then the induced metric $\g_{\hat{\m} \hat{\n}}$ on the brane
reads
 \be
ds^2_{(5)}= -z^2\left(1-{\dot{{\cal R}}^2{R_0^2 \over f
z^2}}\right)d\s^2+z^2d\vec{x}^2+c_0^2R_0^2 f d\vf^2 ~.
\ee

The continuity of the gauge field, on the other hand, is guaranteed by
 the fact that its only non-vanishing component is $A_\vf$ and $X^\vf$ is $\s$-independent.

Apart from the continuity conditions we have to take into account
the junction conditions for the derivatives of the metric and the
gauge field, which read
 \bea
\{\hat{K}_{\hat{\m}\hat{\n}}\}&=&-{1 \over M^4} t_{\hat{\m} \hat{\n}}^{(br)}~, \label{Kjunction} \\
\{n_{M} {\cal F}^{M}_{~N}\de_{\hat{\k}} X^N  \}&=& - {\d S_{br}
\over \d a^{\hat{\k}}}   ~. \label{Fjunction} \eea We denote $\{
H\}=H^{in} + H^{out}$ the sum of the quantity $H$ from each side
of each brane. The extrinsic curvatures are constructed
 using the normal to the brane $n_M$ which points {\it inwards to the corresponding part of the bulk}
each time (we use the conventions of \cite {CR}). The left hand
sides of the above equations are computed in detail in the
Appendix A. Then the junction conditions for the metric are the
following: The $(\s\s)$ component is
 \be
M^4 \left(3{z' \over z} +{f' \over 2 f} \right){ \sqrt{f} \over
R_0 \sqrt{1- \dot{{\cal R}}^2 {R_0^2 \over f z^2}} }  \left(1  -
{1 \over  \b_+}|\dot{T}| \right) = \r~. \label{ss} \ee The $(ij)$
component is
 \bea
 { R_0 M^4 \over  z^2 \sqrt{f}  \left(1-\dot{\cal R}^2 {R_0^2 \over f z^2}\right)^{3/2} }
 \left[  \ddot{{\cal R}}- \b_+ \dot{T}^2\left({ \dot{{\cal R}} \over
 |\dot{T}|} \right)^.  ~~~~~~~~~~~~~~~~~~~~~~~~~~~~~~~~~~~~\right. \nonumber \\
\left.  -2\dot{\cal R}^2  \left(2{z' \over z} +{f' \over 2 f}
\right)(1 -  \b_+ |\dot{T}|)  + {f z^2 \over R_0^2}  \left(3{z'
\over z} +{f' \over 2 f} \right) \left( 1 -{ 1 \over
\b_+}|\dot{T}|^3 \right)  \right]=  -P ~.\label{ij} \eea The $(\vf
\vf)$ component is
 \bea
 { R_0 M^4 \over z^2 \sqrt{f} \left(1-{\dot{{\cal R}}^2{R_0^2 \over f z^2}}\right)^{3/2} }
 \left[ \ddot{{\cal R}}- \b_+ \dot{T}^2\left({ \dot{{\cal R}}
  \over |\dot{T}|} \right)^.  ~~~~~~~~~~~~~~~~~~~~~~~~~~~~~~~~~~~~ \right. \nonumber \\
\left.  - \dot{\cal R}^2  \left(5{z' \over z} +{f' \over 2 f}
\right)(1 - \b_+ |\dot{T}| )  + {4 f z z' \over R_0^2}  \left( 1
-{1 \over \b_+}|\dot{T}|^3 \right)    \right]=  - \hat{P}~.
\label{ff} \eea Finally, the $(\vf)$ component of the gauge field
junction condition is
 \be
 -{c_0  M^2 S \sqrt{f} \over z^4 \sqrt{1-{\dot{{\cal R}}^2{R_0^2 \over f z^2}}}}
 \left(1 - {1 \over \b_+}|\dot{T}| \right)= \hat{L}~, \label{gg}
\ee while its other components dictate that the couplings of the
brane matter  to the other  bulk gauge components vanish \be
l=L=0~. \ee

Equation \reef{ss} gives a Friedmann equation for the brane coordinate ${\cal R}$,
 while equation \reef{ij} gives an acceleration equation. On the other hand,
  \reef{ff} and \reef{gg} are constraint equations for the matter on the
  brane.
   We will for the moment assume that the brane matter is such that $\hat{P}$
    and $\hat{L}$ are given by the latter equations. In a concrete model of matter on the brane the
    latter two equations will introduce some additional constraint for its
    evolution. It is easy to see from \reef{ff} and \reef{gg}, that in  the static limit
      $\hat{P}$ and  $\hat{L}$ tend to zero.

\section{The  {\bf cosmological }dynamics of the $4$-brane}

For discussing the dynamics of the moving brane, we must study the
Friedmann \reef{ss} and the acceleration \reef{ij} equations.
Firstly, in order to bring the Friedmann equation to a more
standard form, we rewrite the metric in the form \be
 ds^2_{(5)}=
-d\t^2+a^2(\t) d\vec{x}^2+ b^2(\t) d\vf^2 ~,
\ee
 with $a= z({\cal
R}(\t))$ and  $b= c_0 R_0 \sqrt{f({\cal R}(\t))}$. The brane
proper time is given by
 \be
\dot{\t}^2=z^2\left(1-{\dot{{\cal R}}^2{R_0^2 \over f
z^2}}\right)~. \label{proper} \ee From now on we will assume
without loss of generality that $\dot{\t}>0$.  It is evident from
the above, that cosmological evolution from mere motion of the
brane in the static bulk is possible only when there is warping in
the bulk. In the unwarped version of this model \cite{PST}, mirage
cosmology is impossible and some bulk time-dependence is
compulsory.

The Hubble parameters for the two scale factors are given by
 \be
H_a \equiv {1 \over a}{ d a \over d \t } =  {z' \over z^2} {
\dot{{\cal R}} \over \sqrt{ 1-{\dot{{\cal R}}^2{R_0^2 \over f
z^2}}  } }~, \label{Ha} \ee
 and
 \be H_b \equiv {1 \over b}{ d b \over d \t } =  {f' \over 2 f z} {
\dot{{\cal R}} \over \sqrt{ 1-{\dot{{\cal R}}^2{R_0^2 \over f
z^2}}  } }~. \ee Then the  ratio of $H_a$ and $H_b$ gives a
precise relation between them for our particular model. It is
given by
 \be
H_b={z f' \over 2 f
z'} H_a~,
\ee
and we notice that since in the model we study it is always  $f' <0$, in the
neighborhood of $r=1$, the two Hubble parameters have opposite sign. This means
 that if the four dimensional space expands, the internal space shrinks.

From \reef{Ha} and \reef{T}, we can express the brane velocity
 $\dot{{\cal R}}$ and the time embedding  $\dot{T}$ as a function of
the Hubble parameter $H_a$ as
 \be
\dot{{\cal R}}= {z^2 \over z'}{H_a \over \sqrt{ 1+ {\cal A}H_a^2
R_0^2 }} ~~~,~~~|\dot{T}|=\sqrt{1+ \b_+^2{\cal A}H_a^2 R_0^2 \over 1+ {\cal A}H_a^2
R_0^2 } ~,\label{vel} \ee
with ${\cal A}={ z^2 \over fz^{'2}}$ evaluated on the brane.

 Then substituting
$\dot{{\cal R}}$ and $\dot{T}$ from \reef{vel} back
to
 \reef{ss}, we obtain the Friedmann equation as a function of the Hubble parameter
\be {M^4 \over R_0} {\cal C} \left(\sqrt{1+ {\cal A} H_a^2 R_0^2 }
- {1 \over \b_+}\sqrt{1+\b_+^2 {\cal A}H_a^2 R_0^2  } \right) =\r
~, \label{efried} \ee where ${\cal C}=\sqrt{f}\left( 3 {z' \over
z}+{f' \over 2f}\right)$ is evaluated  on the brane. We can solve
the above equation for  $H_a$ and obtain a more familiar form \be
H_a^2 = {1 \over 4M^8 {\cal C}^2 {\cal A}}~\r^2 + {M^8 {\cal C}^2
(1-\b_+^2)^2 \over 4 R_0^4 \b_+^4 {\cal A}}\cdot{1 \over
\r^2}-{1+\b_+^2 \over 2 \b_+^2 R_0^2 {\cal  A}}~. \label{fried}
\ee The above Friedmann has unconventional dependence on the
energy density. In particular the inverse square dependence is
known to occur for motions in backgrounds of asymmetrical warping
\cite{asym}.

The equilibrium point $r_c$ of the system is found if we set
$H_a=0$, which gives the brane energy density without matter \be
\r_0=-{M^4 {\cal C}_c \over R_0 \b_+}(1- \b_+)~, \ee where ${\cal
C}_c$ is the value of ${\cal C}$ at $r_c$  and is the same
appearing  in \reef{rsplit}. Let us note that there exists a
second root of \reef{fried} which is not compatible with
\reef{efried}. [This is because we squared twice \reef{efried} to
take \reef{fried}.] The behaviour of the function ${\cal  C}$ is
that, as $r \to -1$ it limits to ${\cal  C} \to  \infty$ and
monotonically decreases and limits to ${\cal  C} \to  -\infty$ as
$r \to 1$. On the other hand ${\cal A}$, is always positive with
${\cal A} \to \infty$ as $r \to \pm 1$, and ${\cal O}(1)$ in the
intermediate region. In the unwarped limit $\a \to 1$ the
Friedmann equation becomes as expected trivial, \ie $H_a=0$.

Before studying various limits of the above equation, let us
define the effective four dimensional matter energy density
$\r_m^{(4)}$ by averaging over the
azimuthal direction (we assume that $\r_m$ is independent
of $\vf$)
\be
\r_m^{(4)}=\int d\vf \sqrt{g_{\vf \vf}} \r_m = {2
\pi \b_+
\over X_+} R_0 \sqrt{f} \r_m ~,
\ee
with similar definitions for the other 4-brane quantities.

Let us suppose now that initially $-1 \ll {\cal R}(\s) < r_c < 1$,
with $1-r_c \ll 1$. The goal is to find how ${\cal R}(\s)$
behaves.  To recover a four-dimensional Friedmann equation at late
times we can assume that the brane energy density is small in
comparison with the static case energy density, \ie $\r_m^{(4)}
\ll \r_0$,  so we can expand \reef{fried} in powers of
$\r_m^{(4)}$ and obtain the following four dimensional form of the
Friedmann equation \be H_a^2 = {8 \pi \over 3} G_{eff} \r_m^{(4)}
+ \D(a) + {\cal O}(\r_m^{(4)~2})~, \ee where the effective
Newton's constant is \be G_{eff}={3X_+ {\cal C}_c (1-\b_+) \over
32 \pi^2 R_0^2 \b_+^2 M^4 {\cal A}~{\cal C}^2 \sqrt{f}} \left[
{{\cal C}^4 \over {\cal C}_c^4}\left({1 + \b_+ \over 1 -
\b_+}\right)^2 -1\right]~. \ee The quantity $\D(a)$ depends on the
parameters of the bulk and  plays the r\^ole   of the the mirage
matter induced on the brane from the bulk and it is given by \be
\D(a)={ (1- \b_+)^2 \over 4 R_0^2 \b_+^2 {\cal A} } \left[ { {\cal
C}_c^2 \over {\cal C}^2 }+{ {\cal C}^2 \over {\cal C}_c^2
}\left({1 + \b_+ \over 1 - \b_+}\right)^2 -2 { 1+ \b_+^2 \over
(1-\b_+)^2} \right] ~. \ee

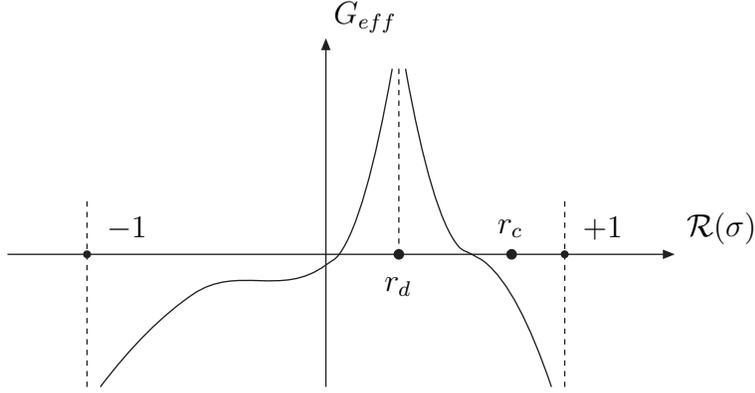
\begin{figure}[t]
\begin{center}
\begin{picture}(200,130)(0,0)

\LongArrow(-40,50)(210,50) \LongArrow(80,0)(80,130)

\Vertex(170,50){1.5} \Vertex(-10,50){1.5}

\Vertex(107.5,50){2} \Vertex(150,50){2}

\DashLine(170,70)(170,0){2} \DashLine(-10,70)(-10,0){2}
\DashLine(107.5,50)(107.5,120){2}

\Curve{(110,120)(130,53)(135,50)(140,47)(165,0)}
\Curve{(-5,0)(30,35)(80,46)(85,50)(105,120)}

\Text(185,60)[c]{$+1$} \Text(5,60)[c]{$-1$}

\Text(230,60)[c]{${\cal R}(\s)$} \Text(150,60)[c]{$r_c$}
\Text(95,140)[c]{$G_{eff}$} \Text(107.5,38)[c]{$r_d$}

\end{picture}
\caption{ The generic form of the effective Newton's constant
$G_{eff}$  as a function of the brane position ${\cal R}(\s)$. As
the brane approaches the equilibrium point $r_c$, we always have
$G_{eff}<0$. Additionally, $G_{eff}$ diverges as $r \to \pm 1$ and
at one point $r_d$ in between.} \label{Geffpic}
\end{center}
\end{figure}

The behaviour of $G_{eff}$ as a function of ${\cal R}(\s)$ is
given  in Fig.~\ref{Geffpic} and has the following important
features: At the points where the geometry becomes conical ($r \to
\pm 1$) the effective Newton's constant is negative and diverging.
In between, there is a point $r_d$, with
 \be z_d = \left({3 (1-\a^8) \over 8 (1-\a^3)} \right)^{1/5}
~, \ee which is a root of ${\cal C}$ and $G_{eff}$ diverges to $+
\infty$. At this point the matter energy density is bound to
vanish. It is important to note that even in the region where
$G_{eff}$ is positive, there is always a strong time variation of
$G_{eff}$, which for \be {1\over G_{eff}}{d G_{eff} \over d\t}=
H_a \d ~, \ee has $\d > {\cal O}(10)$, in contradiction with
observations \cite{obs} which dictate that $\d < 0.1$. But even
more important is the fact that close to the static equilibrium
point, where the cosmology is supposed to mimic best the one of a
codimension-2 brane, we get {\it negative} Newton's constant \be
G_{eff}({\cal R}=r_c)= {3X_+ \over 8 \pi^2 R_0^2 \b_+ (1-\b_+) M^4
{\cal A}~{\cal C}_c \sqrt{f}}   <0~, \ee since we have that for
any value of the parameters and for $r_c$ in the neighborhood of
$r=+1$, it is ${\cal C}_c <0$.

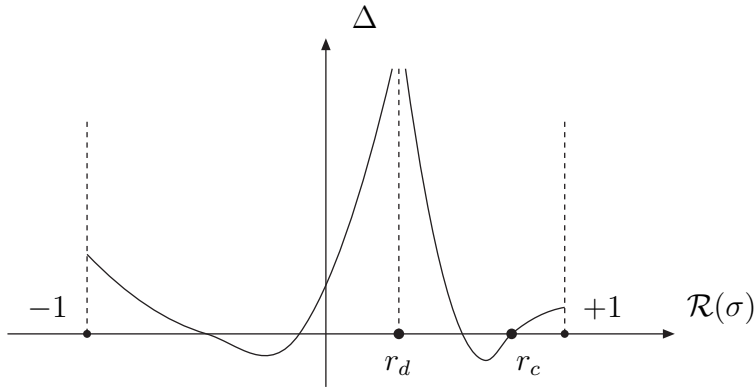
\begin{figure}[t]
\begin{center}
\begin{picture}(200,130)(0,0)

\LongArrow(-40,20)(210,20)
\LongArrow(80,0)(80,130)

\Vertex(170,20){1.5}
\Vertex(-10,20){1.5}

\Vertex(107.5,20){2}
\Vertex(150,20){2}

\DashLine(170,100)(170,20){2}
\DashLine(-10,100)(-10,20){2}
\DashLine(107.5,20)(107.5,120){2}

\Curve{(110,120)(140,10)(150,20)(170,30)}
\Curve{(-10,50)(35,20)(70,20)(105,120)}

\Text(185,30)[c]{$+1$}
\Text(-25,30)[c]{$-1$}

\Text(230,30)[c]{${\cal R}(\s)$}
\Text(157,8)[c]{$r_c$}
\Text(95,140)[c]{$\D$}
\Text(107.5,8)[c]{$r_d$}

\end{picture}
\caption{ The generic form of the mirage matter contribution $\D$
as  a function of the brane position ${\cal R}(\s)$. At the static
equilibrium point $r_c$ it vanishes and at $r_d$ it diverges.}
\label{Dpic}
\end{center}
\end{figure}

Let us also look at the mirage matter contribution $\D$, which  is
depicted in Fig.~\ref{Dpic}. As expected, it vanishes for the
static  equilibrium point $r_c$, it is finite at the boundaries
$r=\pm 1$ and diverges at the root $r_d$ of ${\cal C}$. Again
there is no region in the brane position interval where the
contribution of $\D$ is constant enough to resemble a cosmological
constant contribution to the four dimensional Friedmann equation.

On the opposite limit, that the matter energy density is much
larger  than the static case energy density, \ie $\r_m^{(4)} \gg
\r_0$, we get the expected asymptotics \be H_a^2 = {1 \over 4M^8
{\cal C}^2 {\cal A}}~\r_m^2~, \ee which is a five-dimensional
Friedmann law (with time-varying five-dimensional Newton's
constant) at early times.

Taking under consideration the difficulties of the model to give a
Fiedmann equation with the correct sign of $G_{eff}$, we will not
proceed with the presentation of the analysis of the acceleration
equation \reef{ij}. With this equation, one finds even more
difficulties towards obtaining a realistic four-dimensional
evolution. For example, in the low energy limit, one gets a
coefficient of the linear energy density term, which is not
related to the $G_{eff}$, that we obtained from the Friedmann
equation, in the way it does in standard four-dimensional General
Relativity.

This pathological features of the low energy density limit, where
the  expansion does not appear to have an effective
four-dimensional limit, can find a potential explanation when
looking at the energy continuity equation on the brane. Taking the
covariant divergence of the Israel junction  condition
\reef{Kjunction}, we can obtain the following equation
 \be \nabla^{(4)}_{\hat{\m}}
t^{\hat{\m}~(br)}_{\hat{\n}} = -M^4 \{ \nabla^{(4)}_{\hat{\m}}
\hat{K}^{\hat{\m}}_{\hat{\n}} \}~.
\ee

Then using the Codazzi equation
 \be
\nabla^{(4)}_{\hat{\m}} \hat{K}^{\hat{\m}}_{\hat{\n}}=G_{K
\La}n^{\La} h_N^K \de_{\hat{\n}} X^N ~, \ee and the bulk Einstein
equation, we arrive at the simple expression \cite{CR} \be
\nabla^{(4)}_{\hat{\m}} t^{\hat{\m}~(br)}_{\hat{\n}} = - \{
T^{(B)}_{K \La}h^K_N n^{\La} \de_{\hat{\n}}X^N  \small{\}}~. \ee

Because of the jump of the bulk energy momentum tensor across the
4-brane, the energy-momentum tensor on the 4-brane is not
conserved. This is a usual feature of moving brane cosmologies in
asymmetrically warped backgrounds \cite{asy}. In more detail, the
form of the above equation for the particular model is given by
 \be
{d \r \over d\t}+ 3(\r + P) H_a +(\r + \hat{P}){z f' \over 2 f z'}
H_a = - {S^2 \r H_a \over z' z^7 {\cal C} \sqrt{f}}~, \ee where in
the right hand side we have used the Friedmann equation
\reef{fried}. In a straightforward but lengthy calculation, one
can see that the latter  equation can be derived from \reef{ss},
\reef{ij}, \reef{ff}.  The problems in the four-dimensional limit,
that we faced previously, can be traced to large energy
dissipation off the brane as well as a large work done during the
contraction of the ring-brane.

\section{Conclusions}

In the present paper we made a first step towards the study of the
cosmology of a codimension-2 brane, which is regularized by the
method of lowering its codimensionality (cutting the space close
to the conical tip and replacing it by a ring brane with an
appropriate cap). As a first approximation, we assumed that the
bulk and the cap remain static as the brane moves between them.
The motion of the brane then induces a cosmological evolution for
the matter on the brane. The junction conditions provide the
Friedmann and acceleration equations on the brane.

We can see already from the Friedmann equation that we cannot
recover standard cosmological evolution of the brane at low
energies. The effective Newton's constant is negative  in the
interesting limit that the brane approaches its equilibrium point,
close to the would-be conical singularity. In other words, we
obtain antigravity in this limit. Even away from this point, \ie
when the brane moves away from its equilibrium point, the Newton's
constant varies significantly in contradiction with standard
cosmology. At one position of the internal space, the Newton's
constant even diverges and forces the matter energy density to
vanish. Taking all the above  into account, we did not present the
further analysis of the system by considering the acceleration
equation.

This result is not altered by supersymmetrizing the model
\cite{susy}. The bulk and cap solutions are different from the
non-supersymmetric case due to the presence of the dilaton.
However, the only difference in the Friedmann is a redefinition of
the quantities ${\cal A}$ and ${\cal C}$. In the supersymmetric
case, these quantities read \be {\cal A}_{susy}=4{ z^2 \over
fz^{'2}} ~~~{\rm and}~~~{\cal C}_{susy}=\sqrt{f}\left( {3z' \over
2z}+{f' \over 2f}\right)~.\ee It is easy to see again that the
effective Newton's constant is negative for the motion near the
would-be conical tip and that even away from that point, it is
very strongly varying.

The reason for this unconventional cosmological evolution, is the
use of the specific restricted ansatz for the solution of the
system's equation of motion. The staticity of the bulk was proved
to be an oversimplification. We imagine that this restriction on
the system may result to the appearance of some scalar mode in the
perturbative analysis of  \cite{stabwarp}, which for a certain
region of the brane motion (close to the pole of the internal
manifold) is ghost-like. This mode may then be responsible for the
negative effective Newton's constant. The unrestricted
perturbative analysis in \cite{stabwarp} resulted in a linearized
four-dimensional Einstein equation for distances larger than the
compactification scale. The same happened in \cite{PST} in the
unwarped model perturbation analysis\footnote{The question of
stability of the regularized models was not discussed in
\cite{PST,stabwarp} and it could be that these compactifications
have modes with negative mass squared.} (see \cite{kaloper} for a
related analysis with a brane induced gravity term), even though
it would not have mirage evolution, as we studied it here, because
of the absence of a warp factor. The appearance of the standard
four-dimensional linearized dynamics, shows that they are realized
when the bulk is necessarily time-dependent.

Clearly, the next step should be the study of the system in a setup where the bulk is also time-dependent.
In that respect, brane
cosmology in six dimensions seems to be different for the one in
five dimensions. In Einstein gravity in five dimensions,  one can
always work in a gauge where the bulk is static and the brane
acquires a cosmological evolution by moving into the bulk.
However, in six dimensions there are more degrees of freedom which
make this gauge choice not general. In the present paper, we have
frozen these degrees of freedom hoping to find  consistent
cosmological solutions, but it turned out that this does not give
a viable cosmology. There are several known time-dependent
backgrounds which can be used to look for realistic brane
cosmologies \cite{timedep}. We plan to address this issue in the
near future.

\section*{Addendum}

At the last stages of this work, we became aware of the work of
\cite{md}, where a cosmological evolution, similar in spirit with
the present paper, is discussed. More precisely, the time
evolution of the modulus of a complex scalar field on the brane is
studied, while keeping the bulk static. Our results are in
agreement with \cite{md} and in many respects complementary. In
particular, as in our case, four-dimensional late time cosmology
is not recovered.

\section*{Acknowledgments}

We would like to thank C. Charmousis, J. Gray, D. Langlois and M.
Minamitsuji  for helpful discussions. This work is co-funded by
the European Social Fund and National Resources -(EPEAEK
II)-PYTHAGORAS. E.P is partially supported  by the European Union
through the Marie Curie Research and Training Network UniverseNet
(MRTN-CT-2006-035863).

\def\theequation{A.\arabic{equation}}

\setcounter{equation}{0}

\vskip0.8cm

\noindent

{\Large \bf Appendix A: Computation of extrinsic curvatures}

\vskip0.4cm

 \noindent

In this Appendix we will calculate the extrinsic curvatures on the
brane positions,  which will be used in the main text to evaluate
the junction conditions. Let the brane position in the bulk be
$X^M(\s^{\hat{\m}})$, from which we evaluate the induced metric on
the brane as $\g_{\hat{\m}\hat{\n}}=g_{MN}\de_{\hat{\m}} X^M
\de_{\hat{\n}} X^N $. Firstly, we should calculate the normal
vector of the brane  $n_M$, which is orthogonal to all the tangent
vectors of the brane $\de_{\hat{\n}} X^M$, that is $\de_{\hat{\n}}
X^M n_M =0$, and is  normalized as $n_M n_N g^{MN} =1$.

Once the normal vector is computed, one can evaluate the
projection  tensor $h_{MN}=g_{MN}-n_M n_N$, which is related to
the induced metric as $h^{MN}=\g^{\hat{\m}\hat{\n}}\de_{\hat{\m}}
X^M \de_{\hat{\n}} X^N$ and satisfies also
$\g_{\hat{\m}\hat{\n}}=h_{MN}\de_{\hat{\m}} X^M \de_{\hat{\n}} X^N
$ due to the orthogonality of the normal to the tangent vectors.
Afterwards, the extrinsic curvature  is  given by ${\cal K}_{MN}=
h^{K}_ {M} h^{\La}_ {N} \nabla_K n_\La$ (the covariant derivative
computed with $g_{MN}$) and with trace ${\cal K} = g^{MN} {\cal
K}_{MN}$. The pullback of the  extrinsic curvature on the brane is
given by $K_{\hat{\m}\hat{\n}}={\cal K}_{MN} \de_{\hat{\m}} X^M
\de_{\hat{\n}} X^N$ and has the property that
$K=\g^{\hat{\m}\hat{\n}} K_{\hat{\m}\hat{\n}}={\cal K}$. The
combination which appears in the junction conditions is
$\hat{K}_{\hat{\m}\hat{\n}}=K_{\hat{\m}\hat{\n}}-K
\g_{\hat{\m}\hat{\n}}$.

As is discussed in \cite{ppz}, the metric component $g_{rr}$ in
the radial coordinate direction {\it need not} be continuous. This
then introduces normal vectors $n_M$ for each side of the brane
which are not opposite (since they are normalized with different
$g_{rr}$). Nevertheless, the junction conditions make perfect
sense if the discontinuous $g_{rr}$ and the non-opposite $n_M$'s
are adequately used.

Let us first compute the quantities we are interested in for the
{\it outer bulk section}. The normal vector is \be n^{out}_M=-
{R_0 \over \sqrt{f}}\cdot {1 \over \sqrt{1-{\dot{{\cal R}}^2{R_0^2
\over f z^2}}}} (-\dot{{\cal R}},\vec{0},1,0) \ee

[From now on we suppress the notation ({\it out}) which is to be understood for all subsequent quantities]

The projection tensor  $h_{MN}=g_{MN}-n_M n_N$ has components
\bea
h_0^0={1 \over 1-{\dot{{\cal R}}^2{R_0^2 \over f
z^2}}}~~~,~~~h_0^r=\dot{{\cal R}}h_0^0 ~~&,&~~h_r^0=-\dot{{\cal
R}} {R_0^2 \over f z^2}h_0^0 ~~~,~~~h_r^r=-\dot{{\cal R}}^2 {R_0^2
\over f z^2}h_0^0 \\ h^i_j=\d^i_j~~&,&~~ h^\vf_\vf=1 \eea

The Christoffel symbols for the bulk metric that we need are
\bea
\G^\m_{\n r} = {z' \over z}\d^\m_\n ~~~~&,&~~~~ \G^r_{\m \n} =- {zz'f \over R_0^2}\e_{\m\n}\\
\G^r_{rr}=-{f' \over 2f} ~~~~&,&~~~~ \G^r_{\vf\vf}=-{1 \over
2}c_0^2 ff' \eea

[Also $\G^\vf_{r\vf}=f' /( 2f)$ but we don't need it. All other are zero.]

We can now compute the extrinsic curvatures ${\cal K}_{MN}= h^{K}_
{M} h^{\La}_ {N} \nabla_K n_\La$. Note that since we have
expressed the normal vectors in the brane coordinates, the bulk
derivative is understood to be taken using the projection rule
$h_M^K \de_K = g_{MK} \de_{\hat{\m}}X^K \g^{\hat{\m} \hat{\n}}
\de_{\hat{\n}}$. The calculation gives
\bea
{\cal K}_{00}&=& (h_0^0)^2 n_r[-\ddot{\cal R}-\G^r_{00}+\dot{\cal R}^2(2\G^0_{0r}-\G^r_{rr})]\\
&=&(h_0^0)^2 n_r\left[-\ddot{\cal R} -{zz'f \over R_0^2}+\dot{\cal R}^2 \left(2{z' \over z} +{f' \over 2 f} \right)\right]\\
&&{\cal K}_{05}=-\dot{\cal R}{R_0^2 \over f z^2}{\cal K}_{00}~~~~,~~~~{\cal K}_{55}=\dot{\cal R}^2{R_0^4 \over f^2 z^4}{\cal K}_{00}\\
{\cal K}_{ij}&=&-\G^r_{ij}n_r={zz'f \over R_0^2} n_r \d_{ij}\\
{\cal K}_{\vf\vf}&=&-\G^r_{\vf\vf}n_r={1 \over 2}c_0^2 ff' n_r
\eea

The pullbacks of the extrinsic curvatures  on the brane
$K_{\hat{\m}\hat{\n}}={\cal K}_{MN} \de_{\hat{\m}} X^M
\de_{\hat{\n}} X^N$ are
\bea
K_{\s \s}&=&\left(1-{\dot{{\cal R}}^2{R_0^2 \over f z^2}}\right)^2{\cal K}_{00}=n_r\left[-\ddot{\cal R} -{zz'f \over R_0^2}+\dot{\cal R}^2 \left(2{z' \over z} +{f' \over 2 f} \right)\right] \\
K_{ij}&=&{\cal K}_{ij}={zz'f \over R_0^2} n_r \d_{ij} \\
K_{\vf\vf}&=&{\cal K}_{\vf\vf}={1 \over 2}c_0^2 ff' n_r \eea

Now we can construct the hatted quantities
$\hat{K}_{\hat{\m}\hat{\n}}=K_{\hat{\m}\hat{\n}}-K
\g_{\hat{\m}\hat{\n}}$. They read
\bea
\hat{K}_{\s \s}&=&n_r {f z^2 \over R_0^2}\left(1-{\dot{{\cal R}}^2{R_0^2 \over f z^2}}\right)\left(3{z' \over z} +{f' \over 2 f} \right)\\
\hat{K}_{ij}&=& {n_r \d_{ij} \over   \left(1-{\dot{{\cal R}}^2{R_0^2 \over f z^2}}\right)}   \left[-\ddot{\cal R} +2\dot{\cal R}^2\left(2{z' \over z} +{f' \over 2 f} \right) -{f z^2 \over R_0^2} \left(3{z' \over z} +{f' \over 2 f} \right)\right]  \\
\hat{K}_{\vf\vf}&=&  { n_r c_0^2R_0^2 f \over
z^2\left(1-{\dot{{\cal R}}^2{R_0^2 \over f z^2}}\right) }
\left[-\ddot{\cal R} -4{zz'f \over R_0^2}+\dot{\cal R}^2
\left(5{z' \over z} +{f' \over 2 f} \right)\right] \eea

Let us now compute the quantities we are interested in for the
{\it inner cap section}. The normal vector is \be n^{in}_M= {\b_+
R_0 \over \sqrt{f}}\cdot {1 \over \sqrt{1-\b_+^2{\dot{{\cal R}^2
\over \dot{T}^2}{R_0^2 \over f z^2}}}} \left(-{\dot{{\cal R}}
\over\dot{T}} ,\vec{0},1,0\right) \ee

[From now on we suppress the notation ({\it in}) which is to be understood for all subsequent quantities]

The projection tensor  $h_{MN}=g_{MN}-n_M n_N$ has components
\bea
h_0^0={1 \over 1-\b_+^2{\dot{{\cal R}^2 \over \dot{T}^2}{R_0^2 \over f
z^2}}}~~~,~~~h_0^r= {\dot{{\cal R}} \over \dot{T} } h_0^0
~~&,&~~h_r^0=- \b_+^2{ \dot{{\cal R}}\over \dot{T} } {R_0^2 \over f
z^2}h_0^0 ~~~,~~~h_r^r=-\b_+^2{ \dot{{\cal R}}^2 \over \dot{T}^2} {R_0^2
\over f z^2}h_0^0 \nonumber \\ h^i_j=\d^i_j~~&,&~~ h^\vf_\vf=1 \eea

The Christoffel symbols for the bulk metric that we need are
 \bea
\G^\m_{\n r} = {z' \over z}\d^\m_\n ~~~~&,&~~~~ \G^r_{\m \n} =- {zz'f \over \b_+^2 R_0^2}\e_{\m\n}\\
\G^r_{rr}=-{f' \over 2f} ~~~~&,&~~~~ \G^r_{\vf\vf}=-{1 \over
2}c_+^2 ff' \eea

We can now compute the extrinsic curvatures as
\bea
{\cal K}_{00}&=& (h_0^0)^2 n_r \left[-{1 \over \dot{T}}\left({ \dot{{\cal R}} \over \dot{T}} \right)^.-\G^r_{00}+ { \dot{\cal R}^2  \over \dot{T}^2 }(2\G^0_{0r}-\G^r_{rr})\right]\\
&=&(h_0^0)^2 n_r\left[-{1 \over \dot{T}}\left({ \dot{{\cal R}} \over \dot{T}} \right)^. -{zz'f \over R_+^2}+{ \dot{\cal R}^2  \over \dot{T}^2 } \left(2{z' \over z} +{f' \over 2 f} \right)\right]\\
&&{\cal K}_{05}=-\b_+^2 { \dot{\cal R} \over \dot{T} } {R_0^2 \over f z^2}{\cal K}_{00}~~~~,~~~~{\cal K}_{55}=\b_+^4 {\dot{\cal R}^2 \over \dot{T}^2 } {R_0^4 \over f^2 z^4}{\cal K}_{00}\\
{\cal K}_{ij}&=&-\G^r_{ij}n_r={zz'f \over \b_+^2 R_0^2} n_r \d_{ij}\\
{\cal K}_{\vf\vf}&=&-\G^r_{\vf\vf}n_r={1 \over 2}c_+^2 ff' n_r
\eea

The pullbacks of the extrinsic curvatures  on the brane
$K_{\hat{\m}\hat{\n}}={\cal K}_{MN} \de_{\hat{\m}} X^M
\de_{\hat{\n}} X^N$ are
\bea
K_{\s \s}&=&\dot{T}^2\left(1-\b_+^2{\dot{{\cal R}^2 \over \dot{T}^2}{R_0^2 \over f z^2}}\right)^2{\cal K}_{00}=\dot{T}^2n_r\left[-{1 \over \dot{T}}\left({ \dot{{\cal R}} \over \dot{T}} \right)^. -{zz'f \over \b_+^2 R_0^2}+{\dot{\cal R}^2 \over \dot{T}^2 }\left(2{z' \over z} +{f' \over 2 f} \right)\right] ~~~~~~~~\\
K_{ij}&=&{\cal K}_{ij}={zz'f \over \b_+^2 R_0^2} n_r \d_{ij} \\
K_{\vf\vf}&=&{\cal K}_{\vf\vf}={1 \over 2}c_+^2 ff' n_r \eea

Now we can construct the hatted quantities
$\hat{K}_{\hat{\m}\hat{\n}}=K_{\hat{\m}\hat{\n}}-K
\g_{\hat{\m}\hat{\n}}$. They read
\bea
\hat{K}_{\s \s}&=&n_r {f z^2 \over \b_+^2 R_0^2}\dot{T}^2\left(1-\b_+^2{\dot{{\cal R}^2 \over \dot{T}^2}{R_0^2 \over f z^2}}\right)\left(3{z' \over z} +{f' \over 2 f} \right)\\
\hat{K}_{ij}&=& {n_r \d_{ij} \over   \left(1-\b_+^2{\dot{{\cal R}^2 \over \dot{T}^2}{R_0^2 \over f z^2}}\right)}   \left[-{1 \over \dot{T}}\left({ \dot{{\cal R}} \over \dot{T}} \right)^. +2{\dot{\cal R}^2 \over \dot{T}^2}\left(2{z' \over z} +{f' \over 2 f} \right) -{f z^2 \over \b_+^2 R_0^2} \left(3{z' \over z} +{f' \over 2 f} \right)\right]  ~~~~~ \\
\hat{K}_{\vf\vf}&=&  { n_r c_0^2R_0^2 f \over
z^2\left(1-\b_+^2 {\dot{{\cal R}^2 \over \dot{T}^2}{R_0^2 \over f
z^2}}\right) }  \left[-{1 \over \dot{T}}\left({ \dot{{\cal R}}
\over \dot{T}} \right)^. -4{zz'f \over \b_+^2 R_0^2}+{\dot{\cal R}^2
\over \dot{T}^2 } \left(5{z' \over z} +{f' \over 2 f}
\right)\right] \eea

We can now write down the jumps of the hatted extrinsic
curvatures, which participate in the junction conditions
 \be \{
\hat{K}_{\s \s}  \} = - \left(3{z' \over z} +{f' \over 2 f}
\right) {z^2 \sqrt{f} \over R_0 }\sqrt{1- \dot{{\cal R}}^2 {R_0^2 \over f z^2}}
\left(1 - {1  \over \b_+}|\dot{T}| \right) \ee

\bea
\{ \hat{K}_{i j}  \} =  { R_0  \d_{ij} \over  \sqrt{f}  \left(1-\dot{\cal R}^2 {R_0^2 \over f z^2}\right)^{3/2} } \left[  \ddot{{\cal R}}- \b_+ \dot{T}^2\left({ \dot{{\cal R}} \over |\dot{T}|} \right)^.  -2\dot{\cal R}^2  \left(2{z' \over z} +{f' \over 2 f} \right)(1 - \b_+ |\dot{T}| ) \right. \nonumber \\
\left. + {f z^2 \over R_0^2} \left(3{z' \over z} +{f' \over 2 f}
\right) \left( 1 -{1 \over \b_+}|\dot{T}|^3 \right)  \right]
~~~~~\eea

\bea
\{ \hat{K}_{\vf \vf}  \} = { c_0^2R_0^3 f \over z^2 \sqrt{f} \left(1-{\dot{{\cal R}}^2{R_0^2 \over f z^2}}\right)^{3/2} } \left[ \ddot{{\cal R}}- \b_+ \dot{T}^2\left({ \dot{{\cal R}} \over |\dot{T}|} \right)^. - \dot{\cal R}^2  \left(5{z' \over z} +{f' \over 2 f} \right)(1 - \b_+ |\dot{T}|) \right. \nonumber \\
\left.  + {4 f z z' \over R_0^2}  \left( 1 -{1 \over
\b_+}|\dot{T}|^3 \right)    \right]~~~~~ \eea

Furthermore, the jump of the gauge field is given by
\be \{ n_r
{\cal F}^r_{\phantom{r}\vf} \} = {c_0  M^2 S \sqrt{f} \over z^4
\sqrt{1-{\dot{{\cal R}}^2{R_0^2 \over f z^2}}}} \left(1 - {1 \over \b_+} |\dot{T}| \right) \ee


\begin{thebibliography}{99}








\bibitem{AADD}
  N.~Arkani-Hamed, S.~Dimopoulos and G.~R.~Dvali,
  Phys.\ Lett.\  B {\bf 429} (1998) 263
  [arXiv:hep-ph/9803315];
  I.~Antoniadis, N.~Arkani-Hamed, S.~Dimopoulos and G.~R.~Dvali,
  Phys.\ Lett.\  B {\bf 436} (1998) 257
  [arXiv:hep-ph/9804398];
  N.~Arkani-Hamed, S.~Dimopoulos and G.~R.~Dvali,
  Phys.\ Rev.\  D {\bf 59} (1999) 086004
  [arXiv:hep-ph/9807344].

\bibitem{koz}
  K.~Koyama,
  arXiv:0706.1557 [astro-ph].




\bibitem{CLP}
  J.~W.~Chen, M.~A.~Luty and E.~Ponton,
  JHEP {\bf 0009} (2000) 012
  [arXiv:hep-th/0003067].


\bibitem{5d}
  N.~Arkani-Hamed, S.~Dimopoulos, N.~Kaloper and R.~Sundrum,
  Phys.\ Lett.\  B {\bf 480} (2000) 193
  [arXiv:hep-th/0001197];
S.~Kachru, M.~B.~Schulz and E.~Silverstein,
  Phys.\ Rev.\  D {\bf 62} (2000) 045021
  [arXiv:hep-th/0001206];
S.~Forste, Z.~Lalak, S.~Lavignac and H.~P.~Nilles,
  Phys.\ Lett.\  B {\bf 481} (2000) 360
  [arXiv:hep-th/0002164].
S.~Forste, Z.~Lalak, S.~Lavignac and H.~P.~Nilles,
  JHEP {\bf 0009} (2000) 034
  [arXiv:hep-th/0006139];
C.~Csaki, J.~Erlich and C.~Grojean,
  Nucl.\ Phys.\  B {\bf 604} (2001) 312
  [arXiv:hep-th/0012143];
J.~E.~Kim, B.~Kyae and H.~M.~Lee,
  Nucl.\ Phys.\  B {\bf 613} (2001) 306
  [arXiv:hep-th/0101027].



\bibitem{6dflux}
  S.~M.~Carroll and M.~M.~Guica,
  arXiv:hep-th/0302067;
  I.~Navarro,
  JCAP {\bf 0309} (2003) 004
  [arXiv:hep-th/0302129];
  Y.~Aghababaie, C.~P.~Burgess, S.~L.~Parameswaran and F.~Quevedo,
  Nucl.\ Phys.\  B {\bf 680} (2004) 389
  [arXiv:hep-th/0304256].




\bibitem{susy}
G.~W.~Gibbons, R.~Guven and C.~N.~Pope,
  Phys.\ Lett.\  B {\bf 595} (2004) 498
  [arXiv:hep-th/0307238];
C.~P.~Burgess, F.~Quevedo, G.~Tasinato and I.~Zavala,
  JHEP {\bf 0411} (2004) 069
  [arXiv:hep-th/0408109].






\bibitem{fluxquant}
I.~Navarro,
  Class.\ Quant.\ Grav.\  {\bf 20} (2003) 3603
  [arXiv:hep-th/0305014];
H.~P.~Nilles, A.~Papazoglou and G.~Tasinato,
  Nucl.\ Phys.\  B {\bf 677} (2004) 405
  [arXiv:hep-th/0309042];
H.~M.~Lee,
  Phys.\ Lett.\  B {\bf 587} (2004) 117
  [arXiv:hep-th/0309050].

\bibitem{noncon}
C.~P.~Burgess,
  AIP Conf.\ Proc.\  {\bf 743} (2005) 417
  [arXiv:hep-th/0411140];
N.~Kaloper and D.~Kiley,
  JHEP {\bf 0603} (2006) 077
  [arXiv:hep-th/0601110].







\bibitem{6dsigma}
  A.~Kehagias,
  Phys.\ Lett.\  B {\bf 600} (2004) 133
  [arXiv:hep-th/0406025];
S.~Randjbar-Daemi and V.~A.~Rubakov,
  JHEP {\bf 0410} (2004) 054
  [arXiv:hep-th/0407176];
H.~M.~Lee and A.~Papazoglou,
  Nucl.\ Phys.\  B {\bf 705} (2005) 152
  [arXiv:hep-th/0407208].


\bibitem{Cline}
  J.~M.~Cline, J.~Descheneau, M.~Giovannini and J.~Vinet,
  JHEP {\bf 0306} (2003) 048
  [arXiv:hep-th/0304147].


\bibitem{GB}
  P.~Bostock, R.~Gregory, I.~Navarro and J.~Santiago,
  Phys.\ Rev.\ Lett.\  {\bf 92} (2004) 221601
  [arXiv:hep-th/0311074];
C.~Charmousis and R.~Zegers,
  Phys.\ Rev.\  D {\bf 72} (2005) 064005
  [arXiv:hep-th/0502171].



\bibitem{GBcon}
  E.~Papantonopoulos and A.~Papazoglou,
  JCAP {\bf 0507} (2005) 004
  [arXiv:hep-th/0501112];
G.~Kofinas,
  Phys.\ Lett.\  B {\bf 633} (2006) 141
  [arXiv:hep-th/0506035];
E.~Papantonopoulos and A.~Papazoglou,
  JHEP {\bf 0509} (2005) 012
  [arXiv:hep-th/0507278].











\bibitem{regular}
B.~Carter, R.~A.~Battye and J.~P.~Uzan,
  Commun.\ Math.\ Phys.\  {\bf 235} (2003) 289
  [arXiv:hep-th/0204042];
  M.~Kolanovic, M.~Porrati and J.~W.~Rombouts,
  Phys.\ Rev.\  D {\bf 68} (2003) 064018
  [arXiv:hep-th/0304148];
S.~Kanno and J.~Soda,
  JCAP {\bf 0407} (2004) 002
  [arXiv:hep-th/0404207];
J.~Vinet and J.~M.~Cline,
  Phys.\ Rev.\  D {\bf 70} (2004) 083514
  [arXiv:hep-th/0406141];
I.~Navarro and J.~Santiago,
  JHEP {\bf 0502} (2005) 007
  [arXiv:hep-th/0411250];
J.~Vinet and J.~M.~Cline,
  Phys.\ Rev.\  D {\bf 71} (2005) 064011
  [arXiv:hep-th/0501098].


\bibitem{uzpe}
  P.~Peter, C.~Ringeval and J.~P.~Uzan,
  Phys.\ Rev.\  D {\bf 71} (2005) 104018
  [arXiv:hep-th/0301172].





\bibitem{PST}
  M.~Peloso, L.~Sorbo and G.~Tasinato,
  Phys.\ Rev.\  D {\bf 73} (2006) 104025
  [arXiv:hep-th/0603026].


\bibitem{Gott}
  J.~R.~I.~Gott,
  Astrophys.\ J.\  {\bf 288} (1985) 422.






\bibitem{ppz}
  E.~Papantonopoulos, A.~Papazoglou and V.~Zamarias,
  JHEP {\bf 0703} (2007) 002
  [arXiv:hep-th/0611311].



\bibitem{tas}
  C.~P.~Burgess, D.~Hoover and G.~Tasinato,
  arXiv:0705.3212 [hep-th].

\bibitem{Cuadros-Melgar:2005ex}
  B.~Cuadros-Melgar and E.~Papantonopoulos,
  Phys.\ Rev.\ D {\bf 72}, 064008 (2005)
  [arXiv:hep-th/0502169];
  E.~Papantonopoulos,
  [arXiv:gr-qc/0601011].









\bibitem{mirage}
  A.~Kehagias and E.~Kiritsis,
  JHEP {\bf 9911} (1999) 022
  [arXiv:hep-th/9910174].



\bibitem{5dcosmo}
  P.~Kraus,
  JHEP {\bf 9912} (1999) 011
  [arXiv:hep-th/9910149];
  D.~Ida,
  JHEP {\bf 0009} (2000) 014
  [arXiv:gr-qc/9912002].

\bibitem{equivalence}
  S.~Mukohyama, T.~Shiromizu and K.~i.~Maeda,
  Phys.\ Rev.\  D {\bf 62} (2000) 024028
  [Erratum-ibid.\  D {\bf 63} (2001) 029901]
  [arXiv:hep-th/9912287].




\bibitem{spontan}
  P.~G.~O.~Freund and M.~A.~Rubin,
  Phys.\ Lett.\  B {\bf 97} (1980) 233;
S.~Randjbar-Daemi, A.~Salam and J.~A.~Strathdee,
  Nucl.\ Phys.\  B {\bf 214} (1983) 491.





\bibitem{japs}
  S.~Mukohyama, Y.~Sendouda, H.~Yoshiguchi and S.~Kinoshita,
  JCAP {\bf 0507} (2005) 013
  [arXiv:hep-th/0506050];
H.~Yoshiguchi, S.~Mukohyama, Y.~Sendouda and S.~Kinoshita,
  JCAP {\bf 0603} (2006) 018
  [arXiv:hep-th/0512212].


\bibitem{CR}
  H.~A.~Chamblin and H.~S.~Reall,
  Nucl.\ Phys.\  B {\bf 562} (1999) 133
  [arXiv:hep-th/9903225].


\bibitem{asym}
  D.~Ida,
  JHEP {\bf 0009} (2000) 014
  [arXiv:gr-qc/9912002].



\bibitem{obs}
  O.~G.~Benvenuto, E.~Garcia-Berro and J.~Isern,
  Phys.\ Rev.\  D {\bf 69} (2004) 082002.





\bibitem{asy}
R.~A.~Battye, B.~Carter, A.~Mennim and J.~P.~Uzan,
  Phys.\ Rev.\  D {\bf 64} (2001) 124007
  [arXiv:hep-th/0105091];
 D.~Yamauchi and M.~Sasaki,
  arXiv:0705.2443 [gr-qc].






\bibitem{stabwarp}
  T.~Kobayashi and M.~Minamitsuji,
  Phys.\ Rev.\  D {\bf 75} (2007) 104013
  [arXiv:hep-th/0703029].


\bibitem{kaloper}
  N.~Kaloper and D.~Kiley,
  JHEP {\bf 0705} (2007) 045
  [arXiv:hep-th/0703190].







\bibitem{timedep}
K.~i.~Maeda and H.~Nishino,
  Phys.\ Lett.\  B {\bf 154} (1985) 358;
K.~i.~Maeda and H.~Nishino,
  Phys.\ Lett.\  B {\bf 158} (1985) 381;
A.~J.~Tolley, C.~P.~Burgess, C.~de Rham and D.~Hoover,
  New J.\ Phys.\  {\bf 8} (2006) 324
  [arXiv:hep-th/0608083];
B.~Himmetoglu and M.~Peloso,
  Nucl.\ Phys.\  B {\bf 773} (2007) 84
  [arXiv:hep-th/0612140];
T.~Kobayashi and M.~Minamitsuji,
  arXiv:0705.3500 [hep-th];
 E.~J.~Copeland and O.~Seto,
  arXiv:0705.4169 [hep-th].






\bibitem{md}
  M.~Minamitsuji and D.~Langlois, arXiv:0707.1426 [hep-th].


\end{thebibliography}
\end{document}